# Reevaluating Anomalous Electric Fields at the Air-Water Interface: A Surface-Specific Spectroscopic Survey

*Joseph C. Shirley, Zi Xuan Ng, Kuo-Yang Chiang, Yuki Nagata, Yair Litman, Arsh S. Hazrah, Mischa Bonn*

**Abstract**

The notion that large electric fields at the air-water interface catalyze spontaneous chemical reactions has sparked significant debate, with far reaching implications for atmospheric chemistry and interfacial reactivity. Using vibrational sum frequency generation spectroscopy, we test this hypothesis by directly comparing local electric field strengths at the air–water interface and in bulk water. By applying established vibrational frequency-to-field mappings to the OH stretch of interfacial and bulk water, we extract effective electric field distributions under ambient conditions. Contrary to prevailing claims, our results reveal no spectroscopic evidence for exceptionally strong or long-lived interfacial electric fields. Instead, bulk water consistently exhibits broader distributions and statistically larger field magnitudes. The absence of key spectral signatures, such as redshifted continua, or slowed spectral diffusion, further undermines the idea of anomalous surface fields. Our findings call into question the growing narrative that electrostatic forces at pristine and charge neutral water surfaces can drive chemical reactions and instead highlight the importance of rigorous spectroscopic benchmarks when evaluating interfacial phenomena.

**Introduction**

Interfaces between gas and liquid, liquid and liquid, or solid and liquid are chemically distinct environments, where reactions unfold with unique dynamics and selectivity absent in the bulk.[1–5] Interfaces provide unique physicochemical environments characterized by broken symmetry,[6] altered solvation,[7] and depth-dependent ion distributions,[8] all of which can influence molecular orientation,[9,10] reaction pathways,[11,12] and other interfacial properties.[13,14]

The gas–liquid interface, especially the air–water boundary, has been the focus of intense investigation due to its pivotal role in both natural and engineered systems. In atmospheric chemistry, reactions at aqueous aerosol or cloud droplet interfaces are central to processes like ozone degradation,[15] sulfate formation,[16] and organic aerosol aging.[17] Remarkably, this interface has also been implicated in atypical reaction pathways, including the spontaneous generation of hydrogen peroxide ($H_2O_2$) in pure water microdroplets, a phenomenon attributed to water autoionization coupled with interfacial effects.[18,19] Observations of accelerated reaction rates and unconventional product formation in aqueous microdroplets, emulsions, and sprays have led to growing speculation that interfaces may catalyze reactions even without added reagents or external catalysts.[20,21] These claims, if substantiated, carry profound implications as they redefine the role of interfaces in environmental and synthetic chemistry. Crucially, many interpretations of these findings rest on the specific hypothesis that large or long-lived electric fields exist at the neat air-water interface, providing an electrostatic driving force capable of stabilizing transition states or polarizing bonds.[22] This distinguishes such claims from other known interfacial enhancements that arise from interfacial changes in solvation structure,[23] rather than electrostatic effects.

In this study, we critically evaluate the hypothesis of large, sustained interfacial electric fields at the neat air-water interface. Using vibrational sum-frequency generation (SFG) spectroscopy, we take advantage of the fact that the OH stretch vibration of interfacial water is a direct reporter of the local electric field through well-established frequency-to-field relationships. This approach allows us to quantitatively map interfacial field strengths from the measured vibrational spectra themselves. If strong electric fields (on the order of 100–1000 MV/cm)[24] were present, they would produce characteristic perturbations of the OH stretch band, such as broad continua, pronounced redshifts, and increased anharmonicities, arising from field-induced changes in water's hydrogen-bonding structure. However, the observed SFG spectra reveal only features consistent with a modestly polarized hydrogen-bonded network, with no signatures of such large fields. Applying the same vibrational frequency-to-field mapping to both interfacial and bulk water responses further shows that large electric fields are statistically more probable in bulk water than at the interface, contradicting assertions of strong, spontaneous, field-driven reactivity at the surface.

## Main text
### Defining Interfacial Electric Fields

To assess the hypothesis of field-driven chemistry at aqueous interfaces, we first define what is meant by "electric field" in this context. The concept spans a range of physical scenarios that are not mutually exclusive (certain types of fields can give rise to others) yet each can lead to distinct spectroscopic consequences. Specifically, we consider:

1. Static E-Fields: A static or equilibrium electric field, depicted as normal to the interface, arising from net charge separation (e.g., surface ion gradients) or collective dipolar alignment of water molecules.

2. Field Distributions: A distribution of local "fields" experienced by O–H bonds due to heterogeneous solvation environments and hydrogen-bond fluctuations, some of which may mimic field-like effects.
3. Fluctuations: A temporal overlap between field-like effects and molecular bond orientations, yielding chemically relevant perturbations. For example, having slower solvent or field fluctuations could enhance coupling to reactive coordinates.

While these scenarios offer plausible mechanisms for interfacial electric fields to influence chemical reactivity, the central hypothesis, that such fields are sufficiently large and persistent to drive reactions, has been met with substantial critique. Specifically, challenges have been raised against both the underlying claim that interfacial electric fields are stronger than those in bulk water, and the broader conclusion that such fields, even if present, are capable of inducing chemical transformations.[25,26] These critiques highlight the need to distinguish between different definitions of field magnitude, spatial extent, and timescale, and to assess their spectroscopic consequences using methods capable of resolving molecular-level behavior. In the following sections, we examine the spectroscopic signatures associated with each of these proposed field scenarios to determine whether comparatively large fields or slower field fluctuations at the air–water interface are consistent with experimental observations.

**Vibrational Sum-Frequency Generation Spectroscopy**

To test whether large electric fields are present at the air-water interface, we primarily employ vibrational SFG spectroscopy, a surface-specific technique that directly probes the orientation and hydrogen-bonding environment of interfacial water molecules. The surface sensitivity of SFG spectroscopy arises from the selection rules governing second-order nonlinear optical processes: in isotropic, centrosymmetric media like bulk water, the second-order susceptibility $\chi^{(2)}$ vanishes, ensuring that only interfacial molecules contribute to the signal.[27] In addition, $\chi^{(3)}$ contributions measured with the SFG technique provide a bulk analogue for a direct comparison with the surface $\chi^{(2)}$ response.[28]

The local environment sensitivity of SFG is conferred through the potential energy well of the molecule of interest. Generally, oscillators are anharmonic, and vibrational energy levels can be envisioned as lying within an asymmetric potential. This well-established framework describes both the spacing between vibrational states and the bond dissociation energy. Perturbations such as hydrogen bonding or electric fields modulate the curvature and depth of this potential, thereby shifting the energy of the fundamental (0→1) vibrational transition.[29] Bond weakening, whether due to strong hydrogen bonding or the influence of an electric field, typically leads to a redshift in the vibrational frequency. Therefore, the OH stretch serves as an intrinsic reporter of the local field and bonding environment. While Stark shift spectroscopy of non-water probe molecules has been used to probe interfacial fields, it is more straightforward to use field-induced Stark shifts of the OH groups themselves.[29]

Thus, SFG provides an experimental framework that allows direct comparison between the energy states of water molecules at the surface and in the bulk. The approach is schematically summarized in Figure 1. We note that we study extended, planar surfaces, rather than micron-sized droplets, which is justified by the negligible curvature of micron-sized droplets on molecular length scales.

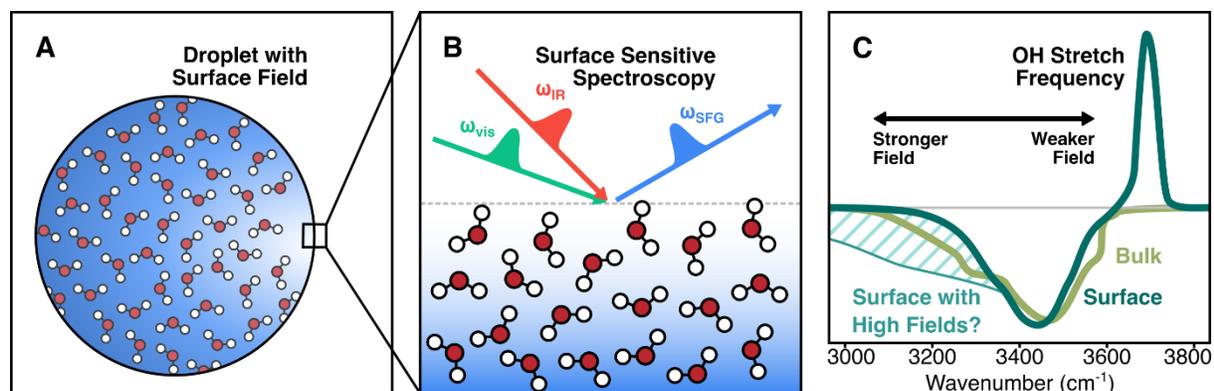

**Figure 1.** (A) Conceptual illustration of a water microdroplet proposed to exhibit a high surface electric field. (B) Schematic of vibrational sum frequency generation spectroscopy, which selectively probes the air–water interface through its sensitivity to non-centrosymmetric regions. (C) Sketch comparing the expected OH stretch SFG responses of bulk and interfacial water. The shaded region highlights a redshifted continuum that would be expected under conditions of comparatively strong static interfacial electric fields.

## Mapping Vibrational Frequencies to Local Fields

While SFG provides access to frequency shifts of the OH stretch vibration, which are connected to electric fields, translating these shifts into electric field strengths quantitatively requires an appropriate model. In this context, vibrational maps serve as a crucial bridge. These empirically derived relationships connect molecular dynamics outputs, such as the instantaneous electric field or electrostatic potential experienced by a molecular bond, to the energy of its vibrational transitions. Accordingly, OH stretch frequency shifts provide a sensitive and quantitative measure of the local interfacial electric field that shapes molecular structure and reactivity. Here, they allow linking experimental observables to the underlying molecular-scale electrostatics.[29] In water, this mapping is particularly powerful because OH stretch frequencies respond strongly to electric fields, both through direct Stark perturbations and through their tight coupling to hydrogen bonding. A hydrogen bond acceptor not only carries a partial negative charge, generating a local field, but also weakens the OH bond of the donor.[30] For water, therefore, it bears repeating that the "electric field" sensed by the OH stretch is not purely electrostatic; the non-Condon effects and other nuclear-coordinate-dependent factors make the inferred field an effective rather than a strictly physical quantity.[30–32] Nevertheless, determining an effective electric field provides a useful comparison with other works that gauge the likelihood of chemical reactivity based upon it. And, given the strong field sensitivity of the OH and OD stretches, vibrational maps have become an essential tool for estimating electric field distributions along hydrogen bonds in aqueous systems.[30,31,33–35]

By applying frequency-to-field mappings to measured spectra, we can estimate the effective electric field experienced by interfacial water molecules. Specifically, we compare the $\chi^{(2)}$ response of the air-water interface with the $\chi^{(3)}$ bulk response, providing a direct measure of field strength distributions in each environment. The $\chi^{(2)}$ signal originates from the non-centrosymmetric interfacial region, which extends approximately 1 nanometer into the liquid, while the $\chi^{(3)}$ spectrum reflects bulk response under externally applied fields, which was obtained by subtracting two experimentally measured Im $\chi^{(2)}$ spectra obtained with phase-resolved SFG spectroscopy. We compare the $\chi^{(3)}$ bulk response with the bulk response inferred from the $\sqrt{(\alpha_{IR} I_{Raman})}$ spectrum, where $\alpha_{IR}$ and $I_{Raman}$ are the bulk infrared and Raman signals, respectively. We use $\sqrt{(\alpha_{IR} I_{Raman})}$ as a bulk reference because the $\chi^{(2)}$ and $\chi^{(3)}$ responses obtained via SFG spectroscopy arise from the product of IR and Raman transitions. Previous studies have shown that this quantity closely reproduces the bulk $\chi^{(3)}$ spectrum at charged interfaces.[28,36,37] To minimize spectral complications from vibrational coupling in neat $H_2O$ or $D_2O$, we analyze the OH stretch of isotopically diluted HOD in $D_2O$ (Figure 2A). The imaginary component of the $\chi^{(2)}$ spectrum reveals two distinct bands: a sharp positive peak near 3700 cm$^{-1}$ originating from the

"free OH" groups at the surface, representing the topmost layer of the surface, and a broad negative band centered around 3400 cm$^{-1}$ from hydrogen-bonded OH groups pointing into the bulk. The $\chi^{(3)}$ spectrum shows a similarly broad feature corresponding to the bonded OH population in the bulk phase. The $\chi^{(2)}$ and $\chi^{(3)}$ spectra used for this analysis were obtained via heterodyne-detected SFG measurements of isotopically diluted HOD in D$_2$O, following established procedures.[11,28] The imaginary component of $\chi^{(2)}$ directly reveals interfacial dipole orientation and hydrogen-bonding character, while the $\chi^{(3)}$ spectrum was extracted from bulk-sensitive difference measurements under applied field conditions (See Methods). The $\alpha_{IR}$ and $I_{Raman}$ spectra were recorded using commercially available instruments as described in the Methods.

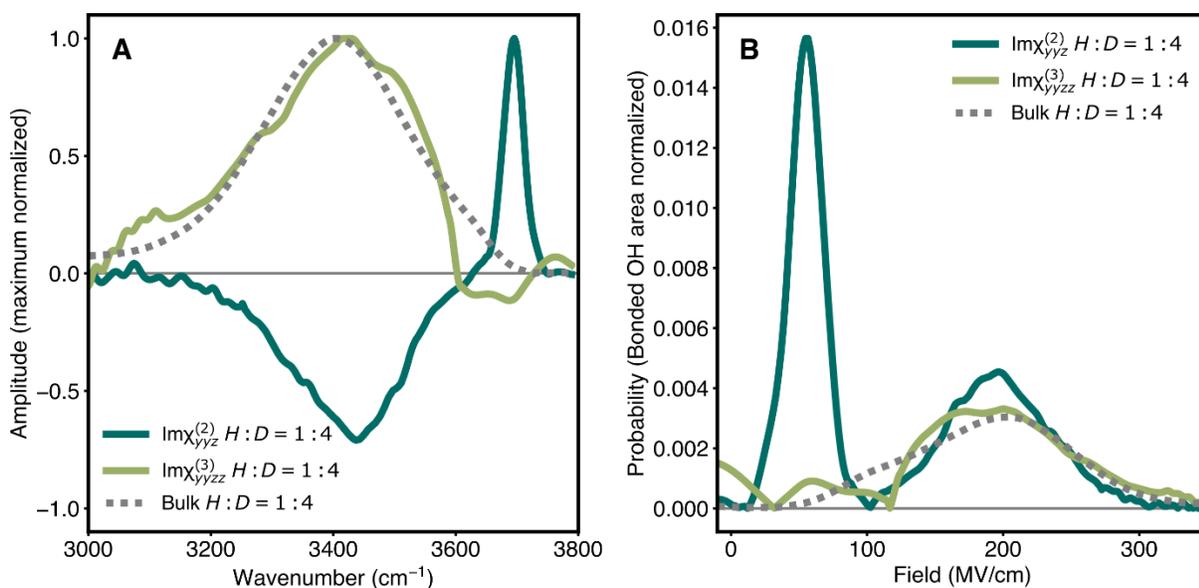

**Figure 2.** A comparison of bulk vs. interfacial field strengths. (A) Imaginary $\chi^{(2)}$ spectrum of the OH stretch in isotopically diluted HDO in D$_2$O at the air–water interface (dark green),[38] compared with the respective $\chi^{(3)}$ spectrum (light green) and $\sqrt{(\alpha_{IR} I_{Raman})}$ spectrum (dotted line), both representing the bulk response. All spectra are normalized to their respective maximum intensities. (B) Probability distribution of electric field strengths projected along the OH bond vector, inferred from the $\chi^{(2)}$, $\chi^{(3)}$, and bulk $\sqrt{(\alpha_{IR} I_{Raman})}$ spectra using established vibrational maps.[39] Distributions are normalized to the bonded OH area of the bulk. The free OH or topmost layer of the air-water

interface experiences relatively small fields, while the OH bonded regions are comparable to those of the bulk. The first moment of the bonded region of the $\chi^{(2)}$ spectrum is 204 MV/cm, 212 MV/cm for the $\chi^{(3)}$ spectrum, and 200 MV/cm for the $\sqrt{(\alpha_{IR}I_{Raman})}$ spectrum.

Using established vibrational maps, we apply the field-to-frequency relationship in reverse to extract the relative probability distribution of electric field strengths along the OH bond from the $\chi^{(2)}$, $\chi^{(3)}$, and $\sqrt{(\alpha_{IR}I_{Raman})}$ spectra (after accounting for relative differences in the response at different frequencies).[39–41] By performing this calculation, one can determine the relative probability of different electric fields along the OH vector in the bulk or at the interface. These results are normalized for the bulk (bonded OH) population and show that the largest electric fields are more likely to occur in bulk water than at the interface (Figure 2B). Molecular dynamics simulations consistently report field strengths in the range of 100-300 MV/cm.[42,43] This directly challenges claims that unusually strong electric fields exist at the air–water interface. If spontaneous water dissociation were driven by field magnitude alone, such reactions would be expected to occur preferentially in the bulk, not at the surface, given the stronger fields in the bulk. Moreover, the broader distribution of field strengths in the bulk suggests a greater diversity or potentially longer-lived energetic states than those accessible at the interface, depending on the source of the broadening. Together, these findings undermine the idea of anomalous surface field distribution (point 2, distributions). There is also no spectral feature (single peak or overall shift) indicative of a narrowly distributed large field (point 1, static field).

To demonstrate this, if comparatively static strong interfacial electric fields were present, one would expect a dramatic alignment of water molecules[44] and enhanced water dissociation,[45] both of which would significantly perturb the SFG spectra.[46] Let's consider in more detail the case of electrolyte solutions containing ions that strongly interact with water. Most inorganic ions lack intrinsic surface activity and preferentially remain solvated in the bulk, as shown by both experimental measurements and molecular dynamics simulations.[47] These ions break inversion symmetry, and give rise to SFG-active water molecules interacting with the ions, even though they are present in the sub-surface region, and not on the surface of water. As shown in Figure 3, ionic solutes such as halides (A) and concentrated hydroxide solutions (B) both lead to dramatic changes in the bonded OH region of the SFG spectrum, including spectral redshifts and broadening below 3400 cm$^{-1}$. These spectral features serve as reliable indicators of perturbations to the local OH environment and heterogeneity, showing that for sufficiently high ionic strength, we can identify the strong hydrogen bonding interaction associated with high field strengths around small ions with a large charge density, such as F$^-$ and OH$^-$. Conversely, iodide is less surface active yet still results in significant changes in the hydrogen-bonded region of the spectrum.[8,48]

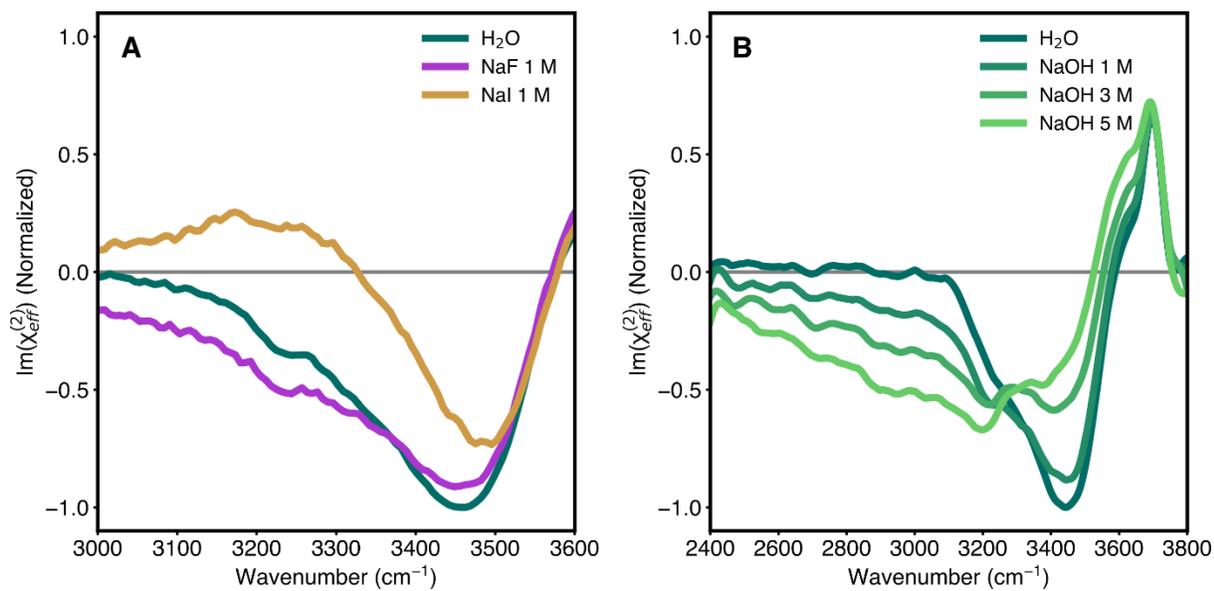

**Figure 3.** (A) The imaginary $\chi^{(2)}$ spectra of neat water and aqueous salt solutions (1M NaF and NaI) at the air-water interface, showing ion-specific perturbations to the bonded OH stretch region. (B) The imaginary $\chi^{(2)}$ spectra of neat water and aqueous NaOH at increasing concentrations. Hydroxide enrichment at the interface leads to progressive broadening and intensity redistribution in the bonded OH region, producing a low-frequency continuum. Adapted from Litman et al.[8]

**Time-resolved Spectroscopy and Long-Lived Fields**

A proposed explanation for enhanced interfacial reactivity is that electric fields at the air–water interface fluctuate more slowly than in bulk, leading to long-lived high-field configurations that could transiently reshape the potential energy surface and promote chemical reactions.[22] While not involving radicals, there is theoretical evidence that rare, transient field conditions are responsible for the autoionization of water.[49] Therefore, multidimensional spectroscopy should be employed to assess the lifetimes of different microscopic structural and energy-redistribution processes. In the following, we consider vibrational and orientational fluctuations. For aqueous bulk and interfaces, two-dimensional infrared (2D IR) and 2D sum-frequency generation (2D SFG) spectroscopy can directly probe the timescales over which vibrational frequencies fluctuate, a process known as spectral diffusion. Specifically, perturbations to the potential energy surface of each probed molecule follow stochastic trajectories. The ensemble average of the autocorrelation functions for each 0-1 transition energy trajectory (Figure 4A) manifests as time-dependent broadening features in the 2D spectra. For water vibrations, spectral fluctuations on the timescale of hundreds of femtoseconds to a couple picoseconds are directly probed (this is limited on the lower bound by instrument resolution and by vibrational lifetime on the upper bound). Fluctuations faster than hundreds of femtoseconds manifest as homogeneous broadening, and fluctuations slower than picoseconds present themselves as quasi-static heterogeneous broadening.[50] When complemented by molecular dynamics simulations that provide sub-femtosecond resolution of local electric field variations, these spectroscopic techniques offer a powerful framework for comparing fluctuation dynamics across bulk and interfacial water.[29]

In bulk water, 2D IR spectroscopy has shown the frequency fluctuation time constant for the OH stretch to be on the order of $180 \pm 40$ femtoseconds (this is the time associated with the autocorrelation function of the trajectory of the 0-1 transition energy).[51] Despite some variation in reported values and interpretations, 2D SFG studies by both Bonn and Tahara show that the spectral diffusion of hydrogen-bonded OH modes occurs on similar timescales as the bulk (240 ± 80 fs[52] and a "few hundred" fs[53]).[54] These results (Figure 4A) demonstrate that the local environments of interfacial OH groups evolve no more slowly than in the bulk, countering the argument for slower interfacial field fluctuations (points 2 and 3, distributions and fluctuations).

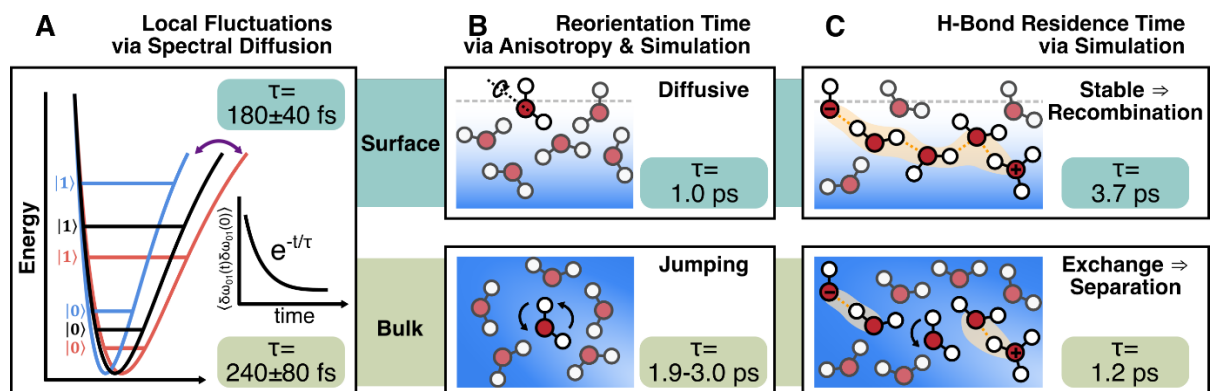

**Figure 4.** Schematic showing time-dependent differences and similarities between the bulk and surface OH stretches. (A) Spectral diffusion timescales, τ, are similar, indicating no change in the rate the potential energy surface is sampled.[51–54] (B) Water reorientation follows different mechanisms at the bulk and the interface, resulting in shorter-lived orientations at the interface.[55–61] (C) The residence time of hydrogen bonds is longer at the interface, potentially leading to a lower probability of charge separation events.[49,62,63]

In the context of orientational fluctuations, which can be related to reaction coordinate alignment, there are two situations to consider. Namely, fluctuations in alignment with a macroscopic laboratory reference frame, which pertain to static fields, and fluctuations in alignment with a molecular reference frame, which pertain to hydrogen bonding arrangements and other local interactions.

Consider first the macroscopic fields, which may be due to fixed charge gradients or external forces. Dipolar reorientation in this reference frame has been characterized via polarization-dependent anisotropy measurements at the surface and the bulk. Time-resolved SFG coupled with molecular dynamics has shown that surface water reorients diffusively with a timescale of around 1.0 ps.[55,56] In contrast, bulk water has been shown to reorient via a jump mechanism that has a corresponding timescale of 2.2 to 2.5 ps from simulation and 1.9 to 3.0 ps from time-resolved IR spectroscopy.[57–61] (Figure 4B) These data indicate that any molecular alignment with macroscopic fixed fields will be longer-lived in the bulk (points 1 and 3, static e-fields and fluctuations).

Now consider the molecular reference frame; geometric alignment of OH bonds in this frame play a critical role in field-mediated reactivity. For example, the autoionization of water requires a precise sequence of events for the process to occur, which involves the generation of a contact ion pair by rare long-range electrostatic solvent fluctuations, followed by reversible proton transfer through the Grotthuss mechanism (10s of femtoseconds), and finally, stabilization of the separated pair by the breaking of a bridging hydrogen-bond "wire" (picoseconds).[49,62] Once the ion pair is generated, the slowest and potentially rate limiting step is the rearrangement of the hydrogen bond network, which enables sustained ion separation. Simulations show that the residence time of hydrogen-bonded OH groups is roughly three times longer at the interface than in the bulk (3.7 vs 1.2 ps),[63] suggesting that recombination of transient ion pairs is more likely at the surface due to hindered network relaxation (Figure 4C) (points 2 and 3, field distributions and fluctuations).

## Conclusions

The air–water interface is structurally distinct from the bulk, yet claims that it supports exceptionally large or long-lived electric fields are not substantiated by direct spectroscopic evidence of the O-H stretching region. Surface-specific vibrational spectroscopy, particularly vibrational SFG, reveals no spectral features consistent with the presence of strong interfacial fields. When vibrational frequency-to-field mappings are applied, the resulting field distributions show that stronger electric fields are more probable in bulk water than at the interface. Our analysis systematically evaluates the three proposed mechanisms by which interfacial electric fields might drive enhanced chemical reactivity:

1. Static E-Fields: Static fields from surface dipoles or charge separation are incompatible with the lack of ion accumulation at the interface,[8] and there are also no spectral features supportive of a narrowly distributed large field.[38,40,41]

2. Field Distributions: Local field heterogeneity is present at both surface and bulk, but spectroscopic data show that broader field distributions occur in the bulk.

3. Fluctuations: Long-lived field configurations, hypothesized to enable field-aligned reactivity, are contradicted by time-resolved and multidimensional spectroscopy, which demonstrate that the bulk and interface sample energy perturbations occur at a similar rate.[51–54] Additionally, OH orientations, related to geometric alignment, are longer-lived in the bulk due to differences in reorientation mechanisms.[55–59] Furthermore, network reorganization occurs faster in the bulk than at the interface, which should facilitate charge separation events.[49,62,63]

These findings directly challenge the view that interfacial electric fields catalyze reactions in aqueous microdroplets or sprays. The interpretation of interfacial electric field-driven catalysis in microdroplet experiments must therefore be reconsidered in light of the absence of spectroscopic field signatures at water interfaces. While interfacial environments can modulate orientation and solvation, their electric field strength and longevity do not differ markedly from the bulk in a way that would enhance bond-breaking reactivity.

Enhanced chemistry at interfaces does not require invoking new physics to explain. Several mechanisms to describe the interfacial enhancement have been previously discussed in detail previously, including altered solvation structure,[23] confinement effects,[64] and evaporative concentration.[65] In many cases, the observed reactivity likely results from a combination of these factors rather than a single dominant cause. Future studies should prioritize time-resolved and multidimensional spectroscopic measurements that directly probe field fluctuations at interfaces. Moreover, comparisons of chemical reactivity in matched interfacial and bulk systems under controlled conditions will help disentangle true field effects from structural or kinetic influences. Clarifying the limitations of interfacial electric field strength is essential not only for fundamental surface chemistry but also for atmospheric science, microdroplet catalysis, and the design of water-interface-based technologies. Ultimately, we conclude that the hypothesis of large electric fields as a catalytic driver at the charge neutral air–water interface is not supported by spectroscopic data and should be reconsidered in the interpretation of interfacial reactivity.

## Methods
### Application of Vibrational Map

To enable quantitative comparison of electric field distributions from vibrational spectra, the spectral amplitude, $\text{Im}(\chi^{(n)})$, was corrected for the dipole derivative, $\mu'$; polarizability, $\alpha_{10}$; and the displacement of the OH stretch from equilibrium for the 1←0 transition, $x_{10}$. This transformation yields an energy distribution, $P$ (Eq. 1). The distribution is calculated using proportionality because it is renormalized before plotting. The correction factors were derived from correlations reported by Skinner and co-workers (Eq. 3-5).[39–41] These operations were performed across the OH stretching region. Simultaneously, frequency-to-field conversions were performed using the same vibrational electrostatic map employed in the spectral density work (Eq. 2).[39–41] The shown frequency-to-field conversion produces two mathematical solutions due to the quadratic dependence of the OH frequency on electric field strength for the given model. Only the positive field solution was retained, as it corresponds to physically meaningful field orientations. This extraneous field rejection was cross-validated against results from a linear mapping models,[33,66] which are included along with a range of other models the Supporting Information in Figure S1.[30,31,33,39,66] These calculations yield electric fields in atomic units (a.u.), which are then converted to MV/cm.

$$P(E) \propto \frac{\text{Im}(\chi^{(n)})}{\mu'(E) * x_{10}(\omega_{10}) * \alpha_{10}(E, \omega_{10})} \qquad \text{Eq. 1}$$

$$\omega_{10} = 3761.6 \text{ cm}^{-1} + \left(\frac{-5060.4 \text{ cm}^{-1}}{\text{a.u.}} * E\right) + \left(\frac{-86225 \text{ cm}^{-1}}{\text{a.u.}^2} * E^2\right) \qquad \text{Eq. 2}$$

$$\mu'(E) \propto 0.71116 + 75.591 \text{ a.u.}^{-1} * E \qquad \text{Eq. 3}$$

$$x_{10}(\omega_{10}) = 0.1024 \text{ Å} - 0.927 * 10^{-5} \text{ Å} * \text{cm}^{-1} * \omega_{10} \qquad \text{Eq. 4}$$

$$\alpha_{10}(E, \omega_{10}) \propto x_{10}(\omega_{10}) * (1.2142 + 3.6206 \text{ a.u.}^{-1} * E) \qquad \text{Eq. 5}$$

**Spectroscopic Measurements**

Vibrational sum-frequency generation (SFG) spectroscopy was used to probe the OH stretch region of interfacial water. To eliminate non-resonant background contributions and directly access the molecular absorptive response, the imaginary component of the second-order susceptibility, Im $\chi^{(2)}$ was previously measured using heterodyne-detected SFG.[37] In addition to isolating the resonant vibrational features, Im $\chi^{(2)}$ carries orientational information: a negative sign corresponds to OH transition dipoles preferentially oriented into the bulk, while a positive sign indicates dipoles pointing toward the vapor phase. Bulk water responses were obtained by extracting the $\chi^{(3)}$ contribution from difference spectra measured using different ion concentrations, as described previously.[11,28] This $\chi^{(3)}$ signal serves as a field-sensitive analogue of the interfacial $\chi^{(2)}$ response and allows direct comparison between surface and bulk environments.

To minimize spectral complications from intramolecular coupling, measurements were performed on isotopically diluted HOD in D$_2$O. In this system, the OH stretch is spectrally isolated, and its response can be interpreted as arising from two dominant populations. The high-frequency positive feature near 3700 cm$^{-1}$ corresponds to weakly hydrogen-bonded OH groups pointing toward the vapor phase, while the broad negative band centered near 3200–3400 cm$^{-1}$ arises from strongly hydrogen-bonded OH groups oriented toward the liquid. The $\chi^{(3)}$ spectrum exhibits a similarly broad feature corresponding to the same frequency range as the bonded OH population.

The experimental SFG spectroscopy instrument uses a 1030 nm fiber laser source (Pharos, Light Conversion). The source is split into two beams. One beam is spectrally narrowed (7 cm$^{-1}$) and used as the upconversion beam. The second is used to drive a bespoke OPA-DFG system and generate mid-infrared light with a bandwidth (1/e$^2$) of ~800 cm$^{-1}$, yielding spectra from 3000 cm$^{-1}$ to 4000 cm$^{-1}$. The beams are combined collinearly and propagate at an angle of 53°, focusing onto a quartz local oscillator (LO) and then the sample. Between the local oscillator and sample is a 5 mm thick SrTiO$_3$ window used to create a 2 ps delay between the signal and LO response. The interference pattern is measured using an Andor Kymera 328i spectrograph and Andor iDus 420 CCD cooled to -100 °C. More details of the spectroscopic system will be disclosed in a forthcoming publication.

The infrared and Raman spectra of isotopically diluted water were measured using a Bruker A225/Q Platinum ATR accessory with a Bruker Tensor II FTIR spectrometer, and a 532 nm WITec alpha300R Raman microscope equipped with a WITec UHTS 300 spectrometer, respectively.

**Acknowledgements**

   We thank Yedam Lee for being *instrumental* in the collection of Raman and FTIR spectra of isotopically diluted water. We are grateful for the financial support from the Max Planck Society and funding from the European Research Council of the European Union (ERC, n-AQUA, Grant 101071937). ASH gratefully acknowledges the financial support from Marie Sklodowska-Curie Actions Postdoctoral fellowship (Spec4DeSal, Grant 101149512) and Natural Sciences and Engineering Research Council of Canada (NSERC).